\documentclass[lettersize,journal]{IEEEtran}
\usepackage{amsmath,amsfonts}
\usepackage{algorithmic}
\usepackage{algorithm}
\usepackage{array}
\usepackage{float}
\usepackage{svg}

\usepackage[normalem]{ulem}
\usepackage[caption=false,font=normalsize,labelfont=sf,textfont=sf]{subfig}
\usepackage{subcaption}
\usepackage[font=normalsize,labelfont=bf]{caption}
\usepackage{textcomp}
\usepackage{stfloats}
\usepackage{url}
\usepackage{xspace}
\usepackage{verbatim}
\usepackage{graphicx}
\usepackage{cite}
\usepackage{caption}
\usepackage{etoolbox}
\usepackage[commandnameprefix=always]{changes}
\usepackage{orcidlink}
\usepackage{hyperref}
\usepackage{xcolor}
\definecolor{darkgreen}{RGB}{0,100,0}

\newcommand{\name}{{\small\textit{AIMNET}}\xspace}
\makeatletter
\newcommand{\ssymbol}[1]{^{\@fnsymbol{#1}}}
\makeatother

\begin{document}

\title{AIMNET: An IoT-Empowered Digital Twin for Continuous Gas Emission Monitoring and Early Hazard Detection}

\author{
    Zifan Zhou$^{1}$\textsuperscript{,*}, 
    Xuan Wang$^1$\textsuperscript{,*}, 
    Yang Yan$^2$, 
    Lkhanaajav Mijiddorj$^2$, 
    Yu Ding$^3$, 
    Tyler Beringer$^2$, 
    Parisa Masnadi Khiabani\,\orcidlink{0009-0002-8569-7224}$^4$, 
    Wolfgang G. Jentner\,\orcidlink{0000-0003-1045-6020}$^4$, 
    Xiao-Ming Hu$^{3,7}$, 
    Chenghao Wang\,\orcidlink{0000-0001-8846-4130}$^{3,6}$, 
    Bryan M. Carroll$^5$, 
    Ming Xue$^{3,5}$, 
    David Ebert\,\orcidlink{0000-0001-6177-1296}$^2$~\IEEEmembership{Fellow,~IEEE}, 
    Bin Li$^1$~\IEEEmembership{Senior Member,~IEEE}, 
    Binbin Weng\,\orcidlink{0000-0002-5706-5345}$^{2,}$$\ssymbol{2}$~\IEEEmembership{Senior Member,~IEEE}

\thanks{*Zifan Zhou and Xuan Wang contributed equally to this work.}
\thanks{$\ssymbol{2}$Binbin Weng (binbinweng@ou.edu) is the corresponding author.}
\thanks{
    \textsuperscript{1}Department of Electrical Engineering, Pennsylvania State University, University Park, PA, USA.
    \textsuperscript{2}School of Electrical and Computer Engineering, University of Oklahoma, Norman, OK, USA.
    \textsuperscript{3}School of Meteorology, University of Oklahoma, Norman, OK, USA.
    \textsuperscript{4}Data Institute for Societal Challenges, University of Oklahoma, Norman, USA.
    \textsuperscript{5}National Weather Center, University of Oklahoma, University of Oklahoma, Norman, OK, USA.
    \textsuperscript{6}Department of Geography and Environmental Sustainability, University of Oklahoma, Norman, OK, USA.
    \textsuperscript{7}Center for Analysis and Prediction of Storms, University of Oklahoma, Norman, OK, USA.
}
}

% The paper headers
\markboth{Submit to IEEE Internet of Things Magazine}%
{Shell \MakeLowercase{\textit{et al.}}: A Sample Article Using IEEEtran.cls for IEEE Journals}

\maketitle

\begin{abstract}
    A Digital Twin (DT) framework to enhance carbon-based gas plume monitoring is critical for supporting timely and effective mitigation responses to environmental hazards such as industrial gas leaks, or wildfire outbreaks \textcolor{black}{carrying large carbon emissions}. We present~{\name}, a one-of-a-kind DT framework that integrates a built-in-house Internet of Things (IoT)-based continuous sensing network with a physics-based multi-scale weather-gas transport model, that enables high-resolution and real-time simulation and detection of carbon gas emissions. {\name} features a three-layer system architecture: (i) physical world: custom-built devices for continuous monitoring; (ii) bidirectional information feedback links: intelligent data transmission and reverse control; and (iii) digital twin world: AI-driven analytics for prediction, anomaly detection, and dynamic weather-gas coupled molecule transport modeling. Designed for scalable, energy-efficient deployment in remote environments, \textcolor{black}{{\name} architecture is realized through a small-scale distributed sensing network over an oil and gas production basin. To demonstrate the high-resolution, fast-responding concept, an equivalent mobile-based emission monitoring network was deployed around a wastewater treatment plant that constantly emits methane plumes.} 
    Our preliminary results through which, have successfully captured the methane emission events whose dynamics have been further resolved by the tiered model simulations. This work supports our position that~{\name} provides a promising DT framework for reliable, real-time monitoring and predictive risk assessment. In the end, we also discuss key implementation challenges and outline future directions for advancing such a new DT framework for translation deployment.
\end{abstract}

\section{Introduction}

Carbon gases, including carbon dioxide (CO$_2$) and methane (CH$_4$), are increasingly valued due to their wide-ranging impacts on climate, human health, and industrial safety management. Environmentally, carbon gases exacerbate global warming, making it a significant contributor to climate change \cite{hansen2007climate}. From a health standpoint, elevated CO$_2$ concentrations in closed or poorly ventilated environments can displace oxygen, leading to symptoms such as dizziness, fatigue, and cognitive impairment~\cite{jacobson2019direct}. \textcolor{black}{In addition, CH$_4$ emissions from oil and gas production processes are often accompanied by various volatile organic compounds (VOCs), which are associated with serious health risks, including cancer\cite{agarwal2016vocc}}. From the industrial aspect, CH$_4$ is highly flammable, posing severe risks in mining, and oil and gas production sectors\cite{alvarez2018assessment}. These challenges highlight the important needs for a scalable real-time and high-resolution gas leak monitoring systems to facilitate timely and effective actions to mitigate environmental harm, safeguard public health, prevent catastrophic accidents and enhance economical robustness.

Recent advances in the Internet of Things (IoT) are ushering in a new era of emission monitoring, enabling real-time, automated data acquisition with cost-effective and remote deployment~\cite{zhang2023recent}. \textcolor{black}{Industries have widely adopted these systems for specific, critical tasks like gas monitoring on production sites, structural health monitoring of bridges, and water-level monitoring for flood alerts.}
However, the practical utility of current IoT-based monitoring systems has been constrained by a critical limitation: narrow geographical coverage. Most existing deployments are confined to specific facilities or urban areas~\cite{hu2025observation}, limiting their ability to capture the broader spatiotemporal dynamics of gas transport and accumulation. This limitation hinders the accurate attribution of emissions sources, environmental impact assessment, and the development of reliable forecasting models, thereby undermining large-scale regulation and mitigation strategies. 

To address these limitations, researchers have begun developing large-scale IoT platforms aimed at expanding spatial coverage and enhancing monitoring capabilities across broader regions, such as IoT-Mobair for air quality tracking~\cite{dhingra2019internet} and the Internet of Maritime Things for marine surveillance~\cite{song2021internet}. These systems leverage low-power wide-area networks (LPWANs) and heterogeneous sensor arrays to extend coverage across urban, industrial, and remote environments. However, despite their promise, critical limitations remain. First, network reliability is a persistent challenge: LPWAN links often experience intermittent outages in rugged terrain or under extreme environmental conditions, leading to delayed or lost data. Second, spatial and temporal resolution is frequently insufficient, with many systems operating at coarse scales (e.g., $\geq 4.0$ km or $\geq 1.0$ h~\cite{epa58d}), failing to capture transient or localized emission events. Third, intelligence and responsiveness are limited, as most platforms function primarily as passive data collectors without integrated modeling, anomaly detection, or forecasting capabilities. Finally, data visualization is often rudimentary, lacking context-aware synthesis or decision-support interfaces such as dynamic heatmaps or multi-parameter correlation views, which are crucial for rapid interpretation and response in multi-hazard scenarios.

Addressing these challenges requires a paradigm shift from passive, static data collection toward an intelligent environmental monitoring paradigm. Unlike conventional IoT systems that rely on fixed sensor placements, we can leverage mobility and autonomy to enable dynamic data collection. The system can actively navigate the environment, detect elevated emission zones, and reposition itself in real-time for more accurate and localized monitoring of hazardous gases \textcolor{black}{or potentially hazardous leak zones}.
To enable such autonomous functionality, the system must possess a comprehensive understanding \textcolor{black}{and predictive simulation} of its deployment environment. 
Digital Twin (DT) provides this capability by establishing a real-time, data-driven digital twin world (DTW) of the physical world. 
\textcolor{black}{Unlike traditional simulation, the DTW mirrors the dynamics of the real physical system in real time, enabling intelligent decision-making based on both historical data and the current system state. Additionally, the DTW can predict future phenomena through a comprehensive understanding of system dynamics, including “what-if” scenarios generated by its own decision-making capabilities.}
As an emerging paradigm, DT offers a powerful \textcolor{black}{technical} framework for bridging the gap between the physical world and DTW in many practical applications.
For example, DT has been used in urban traffic management to support predictive simulations for real-time decisions like adjusting traffic light logic~\cite{xu2023smart}.  
Beyond transportation, DT also finds applications in smart buildings (e.g., energy optimization, temperature control), precision manufacturing (e.g., predictive maintenance, performance monitoring)~\cite{N23}, \textcolor{black}{and some human-centric systems~\cite{chen2023networking}}.
However, the integration of DT into intelligent environmental monitoring, particularly for carbon-based gas emission detection, remains unexplored.
\textcolor{black}{To the best of our knowledge, our work is also the first monitoring framework that employs a basin-scale distributed sensing network for detecting methane plumes.}
By fusing IoT-based sensing with AI-driven analytics and physics-informed simulations, a DTW can continuously interpret, predict, and respond to the environment. 
Specifically, \textcolor{black}{this framework enables the integration of the advanced Weather Research and Forecasting model with greenhouse gases (WRF-GHG) modeling, commonly used to predict and retrieve regional, low-resolution greenhouse gas dynamics, and geometry-resolving large-eddy simulation (LES) to resolve methane emissions and identify their sources with unprecedented spatiotemporal resolution and accuracy.} Furthermore, it enables context-aware visualization interfaces that facilitate rapid decision-making in complex scenarios. In essence, DT addresses the core deficiencies of existing monitoring systems, laying the foundation for regional-scale monitoring and early hazard detection.

We introduce~{\name}, an IoT-empowered DT platform designed to deliver intelligent monitoring and real-time decision support. {\name} comprises three core components as introduced in Section~\ref{sec:architecture}. \textcolor{black}{By seamlessly integrating customized hardware, resilient networking, and AI-empowered analytics, {\name} addresses the critical deficiencies of existing systems, offering a robust, scalable foundation for next-generation environmental monitoring.}

The main contributions of this paper are:

$\bullet$ \textbf{Built-In-House IoT Gas Sensing Instrument:} We create a non-dispersive infrared (NDIR) technology-based IoT gas sensor that features a single digit-parts per million (ppm) sensitivity towards CH$_4$, and a highly-integrated hardware and software system allowing the continuous and robust field operation, and self-calibration functions. \textcolor{black}{Experimental results show that our device attains approximately $90\%$ accuracy of the benchmarking LI-7700 instrument in detecting CH$_4$ leaks, while taking only a few percentage of the cost compared to its high price.}

$\bullet$ \textbf{Continuous Gas Emission Monitoring Network:} Leveraging the lightweight Message Queuing Telemetry Transport (MQTT) protocol, \textcolor{black}{we design a bidirectional communication layer that supports real-time data streaming, intelligent sensing control, and remote device management, even in harsh environments.}  % and adaptive duty-cycling to extend battery life 

$\bullet$ \textbf{AI-Powered Modeling, Visualization and Analysis:} Empowered by our DT framework which integrates an intelligent sensing data analysis layer and a physics-based multiscale weather-gas modeling layer, a web-based visualization interface with interactive heat maps and multi-parameter correlation views is created. This visualization platform enables the real-time modeling and visualization of the emission dynamics over targeted infrastructures. 

$\bullet$ \textbf{Real-World Implementation at Scale:} We implement our~{\name} architecture through both a small-scale distributed sensing network with $23$ static field nodes over an oil and gas field, and a mobile-based plume scanning network mimicking the envisioned large-scale distributive network for proving the concept. The static deployment primarily supports large-scale visualization, while data trustworthiness is validated through mobile detection using a high-precision LI-7700 benchmark sensing instrument.

\section{AIMNET Architecture}
\label{sec:architecture}
\begin{figure*}[htbp]
\centering
    \includegraphics[width=1\linewidth]{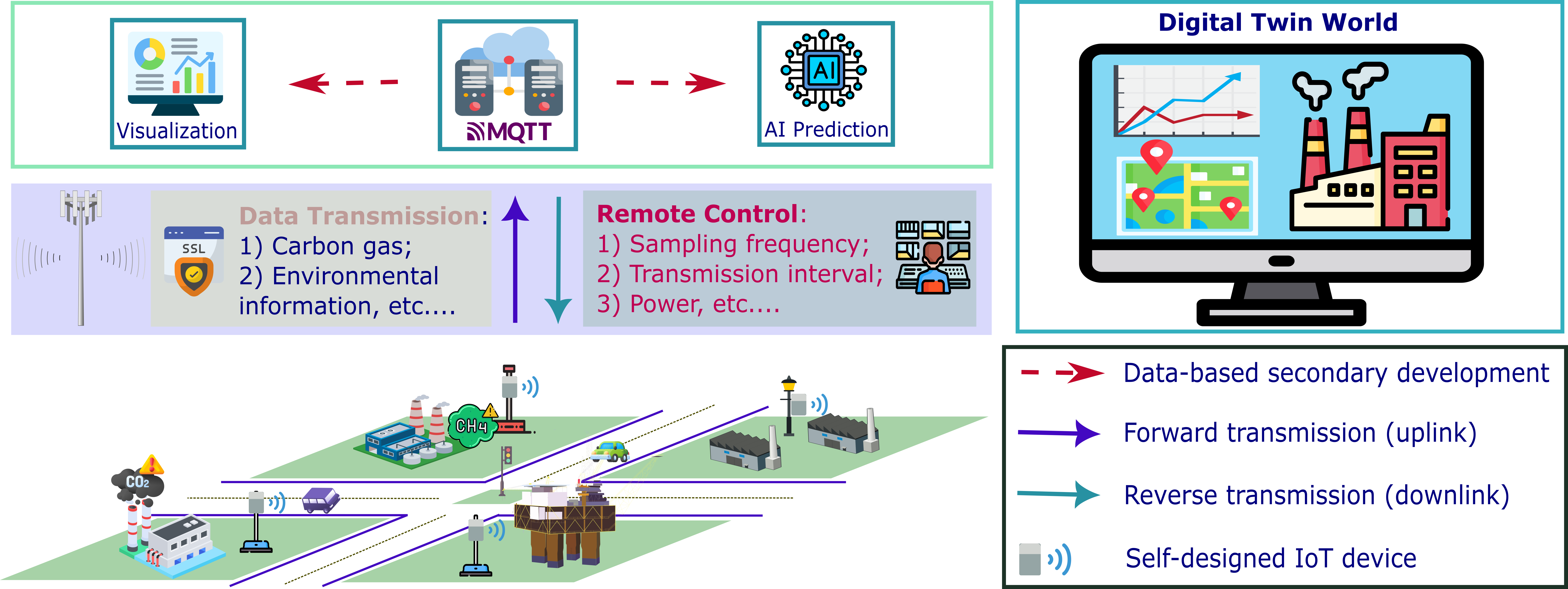}
    \caption{The large-scale Digital Twin gas monitoring framework integrated with a distributed IoT gas sensing network, advanced gas transport modeling and data visualization technologies.}
    \label{fig:MethaneWatch}
\end{figure*}

{\name} is architecturally structured into three primary layers: i) Physical world: front-end data collection layer, ii) Bidirectional information feedback links: middle-end communication layer, and iii) Digital twin world: back-end analytics layer, as illustrated in Fig.~\ref{fig:MethaneWatch}.
The main functions of the physical layer are
environmental data collection and transmission. The device’s microcontroller intelligently coordinates each electronic component for low-power, adaptive, and autonomous field operation.
The communication layer is built upon the lightweight MQTT protocol and supports reliable bidirectional information exchange between field devices and the DTW. 
\textcolor{black}{
It enables the upstream transmission of real-time sensor data to the analytics platform and the downstream delivery of control commands from operators to edge devices.
The DTW functions as a cloud-based platform for data-driven secondary application development and intelligent management.} It integrates an interface for real-time and historical data visualization, along with AI-driven modules for data enhancement, anomaly detection, and predictive modeling. This enables actionable environmental insights and supports the development of advanced applications for hazard detection and decision-making.

\subsection{Physical World: Front-end Data Collection Layer}

\begin{figure}[htbp]
\centering
    \includegraphics[width=1.0\linewidth]{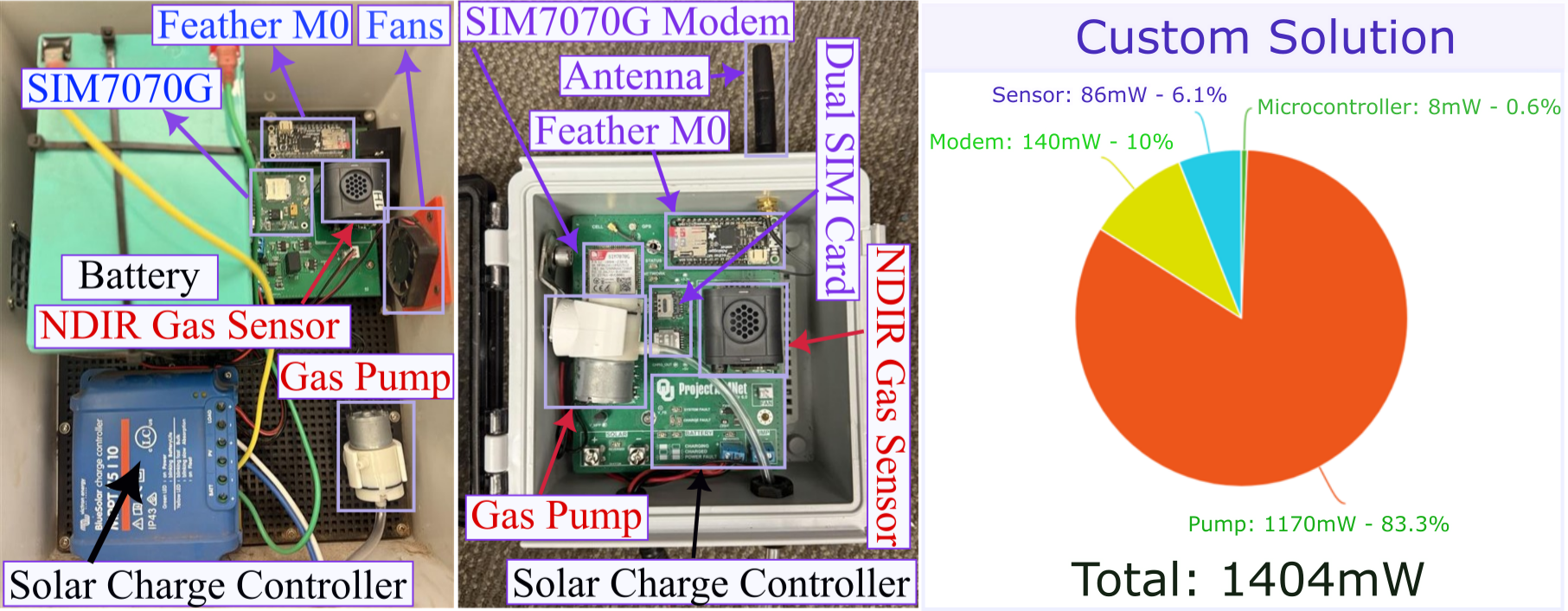}
    \caption{Prototypes of IoT gas sensing instrument: left one features a completed system with battery and power management modules; middle one is the revised version isolating the core sensing control from the battery module for enhanced field operating robustness; right one shows the total power consumption of our device.}
    \label{fig:hardware}
\end{figure}

The physical layer serves as the data source, acquiring and transmitting high-fidelity environmental data from the field. Achieving this requires overcoming significant challenges, including the need for long-term operational reliability, energy efficiency, and data accuracy in harsh, uncontrolled outdoor environments.
\textcolor{black}{To meet these demands, we develop a fully integrated, custom-built IoT sensing device, as illustrated in  Fig.~\ref{fig:hardware}. Unlike conventional commercial gas detectors that rely on off-the-shelf development boards or pre-packaged sensor modules, our approach provides full control over the hardware stack and enhances mechanical durability. We retain only the essential functional components (i.e., sensing front-end, cellular modem, and voltage converter) and consolidate them onto a unified, application-specific PCB. This integration significantly reduces hardware complexity, minimizes idle and peak power consumption, lowers the bill of materials cost, and decreases overall size without sacrificing functionality. As Fig.~\ref{fig:hardware} shows, our device achieves a power consumption of just $1404\text{ }mW$, compared to the LI-7700’s $8-41\text{ }W$.}

\textcolor{black}{The hardware architecture is designed for autonomous field operation. In detail, the control and sensing functions are managed by an Adafruit Feather M0 microcontroller, which serves as the central hub.} For high-fidelity sensing, it utilizes an NDIR technology sensor, chosen explicitly for outdoor detection due to its robust resilience and significantly reduced susceptibility to environmental influences compared to electrochemical sensors. 
Furthermore, this sensor offers long-term reliability with significantly lower power consumption compared to other optical instruments.
A solar-powered energy harvesting module provides the power, ensuring consistent performance even for an entire week without recharging. This enables a low-power, narrow-band SIM7070G communication module to reliably and efficiently transmit data.

A cornerstone of our design is the tight integration of hardware and software to ensure data credibility. While optical NDIR sensors are cost-effective and sensitive, their accuracy can be affected by environmental factors, such as humidity, temperature, and atmospheric pressure~\cite{dong2025environmental}. To mitigate this, we move beyond traditional signal processing and develop a custom, regression-based machine learning (ML) model to refine data as introduced in~\ref{sec:virtualworld}.

\subsection{Bidirectional Information Feedback Links: Middle-end Communication Layer}

\textcolor{black}{The {\name} communication layer, bridging the physical world and DTW, addresses the critical challenge of reliability in regional IoT deployments. It enables reliable data collection from the physical world and supports timely operational suggestions from the DTW to physical entities}. In harsh environments, such as rural oil fields, intermittent connectivity and limited hardware can hinder data transmission. Our architecture provides a resilient and secure framework explicitly engineered for these demanding conditions.

Reliability, ensuring stable data streaming and timely command delivery, is the foundation of DT systems. However, harsh networking conditions in the field often result in high packet loss and delayed command delivery, and commercial IoT platforms can exacerbate the problem due to their lack of transparency and control over critical performance parameters. To overcome these limitations, {\name} uses a self-hosted MQTT broker. \textcolor{black}{We chose MQTT for its efficient publish-subscribe architecture,  which avoids the high latency of HTTP’s request-response model for each data exchange and offers greater reliability than UDP-based protocols like CoAP. This is because MQTT provides robust message acknowledgment and session recovery, which are essential for real-time digital twin synchronization.} This approach provides fine-grained control by assigning each IoT device a unique topic for both publishing sensor data and receiving commands. 
\textcolor{black}{In detail, for forward data transmission, each IoT device is assigned a unique MQTT topic. The central MQTT broker classifies incoming data based on these topics, allowing the DTW to reconstruct the real-world environmental dynamics using the topic-associated data streams. For reverse management, such as delivering operational suggestions or enabling remote configuration of device parameters, IoT devices are designed to receive commands either through designated MQTT topics or via direct operator SMS messages.}

\textcolor{black}{To balance energy efficiency and timeliness, we use a duty-cycled policy, 5-minute sampling followed by 1-minute uplink transmission. We further evaluated the communication performance to verify reliability under real-world conditions. Transport latency (device-to-dashboard once sent) is seconds-scale, while dashboard freshness remains within 6 minutes. With MQTT QoS 1 and session recovery enabled, the system achieves a packet loss below 0.1\% for transmission intervals of $\geq$4 seconds, with each node maintaining an average uplink bandwidth below 1 kbps.}

\subsection{\textcolor{black}{Digital} Twin
World: Back-end Analytics Layer}
\label{sec:virtualworld}
\textcolor{black}{The last layer serves as the intelligence core of~{\name}, functioning as both the data center and the DTW.} It hosts the MQTT broker, manages data storage, supports visualization, and runs AI-driven analytics to enable real-time environmental monitoring, predictive analytics, modeling, and system-wide decision-making.

\subsubsection{AI-empowered Calibration} \textcolor{black}{To enhance data completeness and mitigate the influence of environmental factors on sensor readings, we deploy an ML model trained on a comprehensive dataset comprising measurements from both laboratory and field environments. The dataset includes CH$_4$ and CO$_2$ concentrations along with corresponding environmental parameters, enabling the model to learn correction patterns under diverse conditions. Model performance is validated using two independent datasets: one collected in an indoor environment and the other in an outdoor, dynamic setting. Results indicate that the model reduces the mean absolute deviation in methane readings to approximately $\pm3$ ppm and achieves coefficients of determination of $R^2=0.948$ for the indoor dataset and $R^2=0.908$ for the outdoor dataset, enabling DT to maintain stable performance even under highly variable field conditions.}

\subsubsection{DT-enabled Modeling and Analysis} 
The heart of this layer is a DTW that dynamically fuses data from the physical world and physics-based multiscale models to create a real-time, virtual representation of the monitored environment. 
Specifically, we use the WRF-GHG to simulate near-real-time meteorological conditions and greenhouse gas concentrations at a resolution of approximately 1 km. These outputs then serve as boundary conditions for a geometry-resolving LES, which enables much finer-scale simulations at resolutions ranging from 0.5 to 5 m.
This multiscale representation is continuously updated with live sensor data, accurately reflecting the evolving physical conditions of the region. By the closed-loop synchronization between physical and virtual components, the \textcolor{black}{DTW} recalibrates in response to environmental changes (e.g., wind direction, humidity, temperature), consistent with the concept of ``living DTs"~\cite{elghaish2024predictive}.
The \textcolor{black}{DTW} also integrates simulation and inverse modeling techniques to optimize sensor placement, identify likely emission sources, and evaluate mitigation strategies, which enhances the system's capability to assess gas leaks under varying meteorological conditions.

\subsubsection{Data Visualization} 

\begin{figure}[htbp]
\centering
    \includegraphics[width=1.0\linewidth]{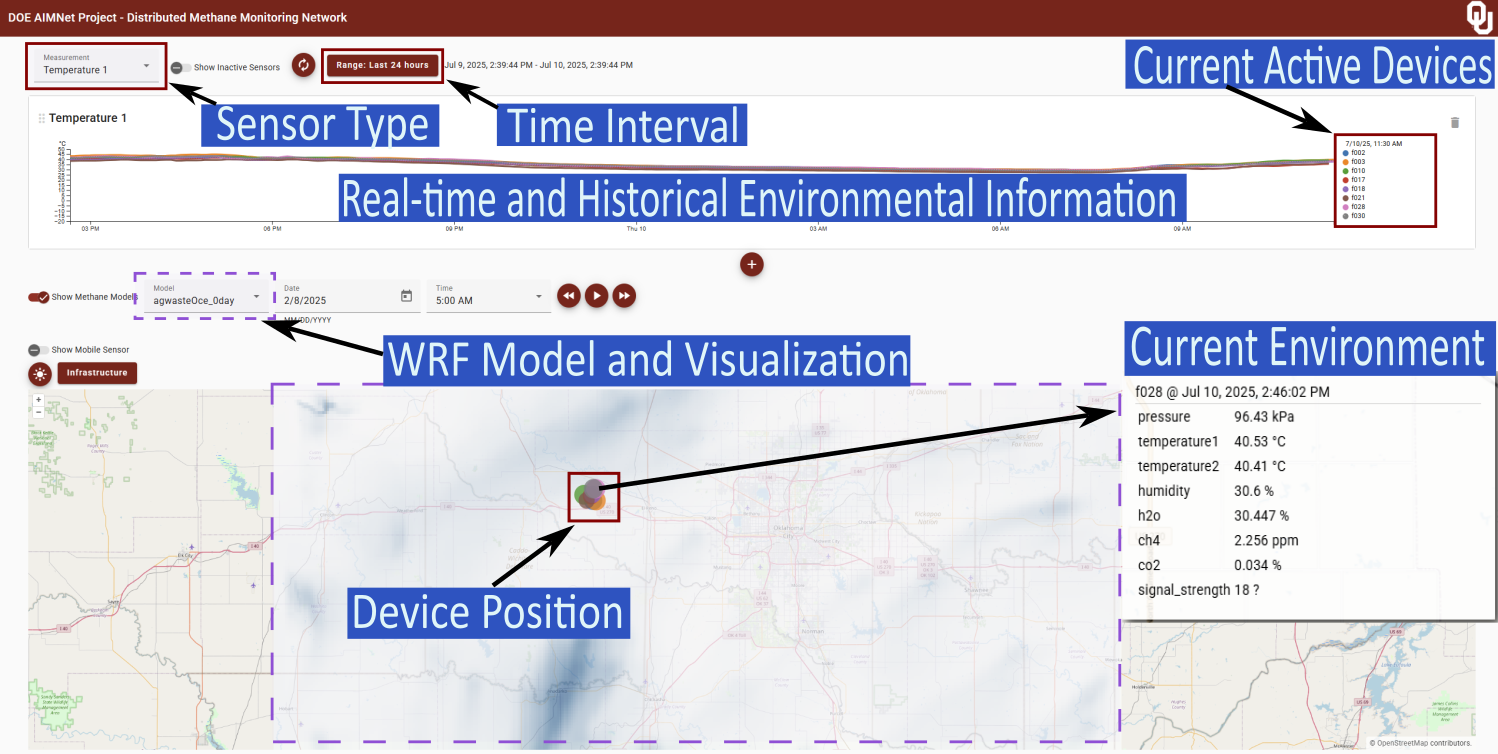}
    \caption{Real-time and historical visualization.}
    \label{fig:web_interface}
\end{figure}

In an era demanding increasingly granular and responsive environmental oversight, traditional data analysis methods often fall short of providing the dynamic insights necessary for effective decision-making. To bridge this critical gap, we develop an interactive environmental monitoring dashboard as depicted in Fig.~\ref{fig:web_interface} that transforms disparate sensor readings into actionable intelligence. 
At its core, the system features a robust backend that supports high-volume data ingestion from diverse sources, including fixed sensors, mobile platforms, and external weather stations, providing a comprehensive understanding of the environment. 
The user-facing interface offers an intuitive experience, integrating real-time geospatial maps with interactive time-series visualizations. This synchronized display enables users to identify subtle trends and anomalies across customizable time windows and immediate localization.
% them in space.
%
Both real-time and historical data are presented through interactive heat maps and multi-parameter correlation views. 
AI modules enhance system intelligence by detecting anomalies and imputing missing data, effectively addressing both short-term gaps and longer outages by leveraging neighboring sensor information. Integrated weather model overlays enrich sensor data with atmospheric context, enabling users to trace potential emissions sources. 
Advanced features, such as auto-refreshing, sensor isolation controls, and optional infrastructure overlays (displaying oil and gas facilities or calibrated methane measurements), further enhance analytical precision.
These capabilities provide a comprehensive and actionable view of the evolving environment, empowering operators to detect hazards early and make proactive decisions with confidence.

\section{Preliminary Implementation and Evaluation}
\begin{figure}[htbp]
\centering
    \includegraphics[width=1.0\linewidth]{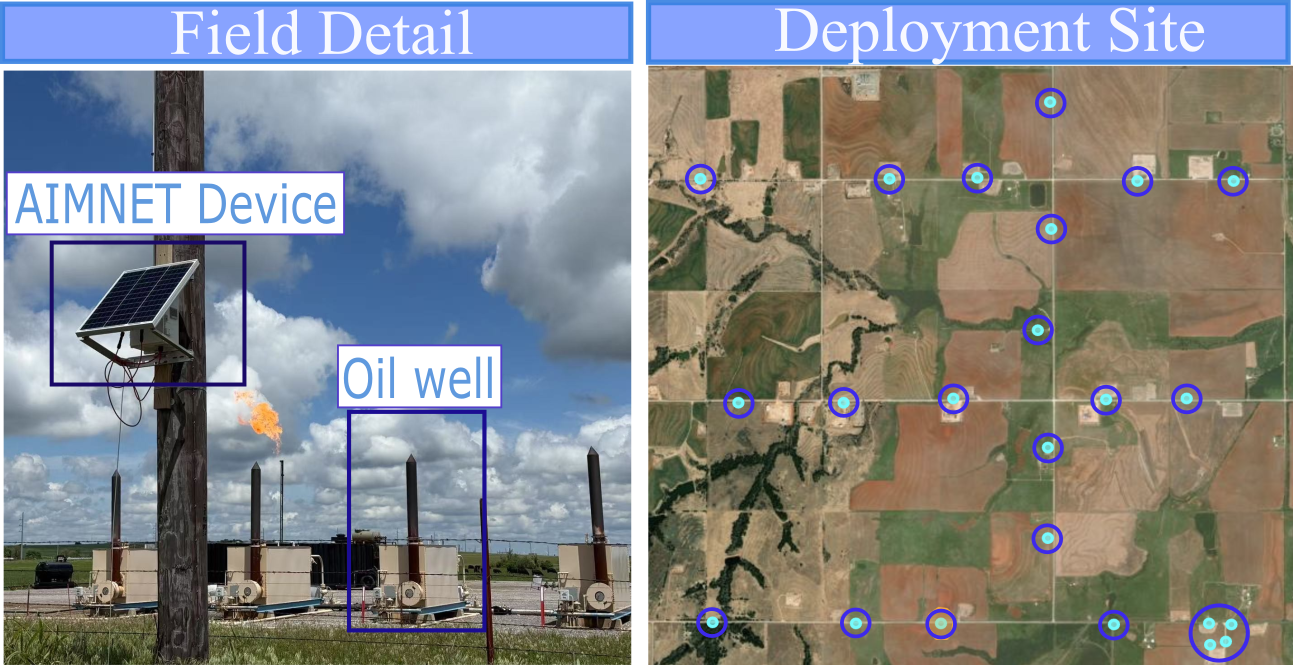}
    \caption{Experimental deployment (Map from Google Maps).}
    \label{fig:implementation}
\end{figure}

To evaluate the trustworthiness of~{\name}, we deploy both static and dynamic operational modes. While both CH$_4$ and CO$_2$ are measurable, the implementation focuses on CH$_4$ due to its greater challenges, though the method also applies to broader CO$_2$ emission monitoring. In the static mode, a total of 23 IoT devices are deployed onto the power posts along the road over the Anadarko basin (Fig.~\ref{fig:implementation}), which includes environmentally complex regions such as oil wells. These devices were installed approximately 2.5 meters above the ground and spaced at intervals of  0.5 miles to optimize both solar energy harvesting and high-resolution gas detection. The static network has been continuously operational from March 2024 to July 2025.

\begin{figure}[htbp]
\centering
    \includegraphics[width=1.0\linewidth]{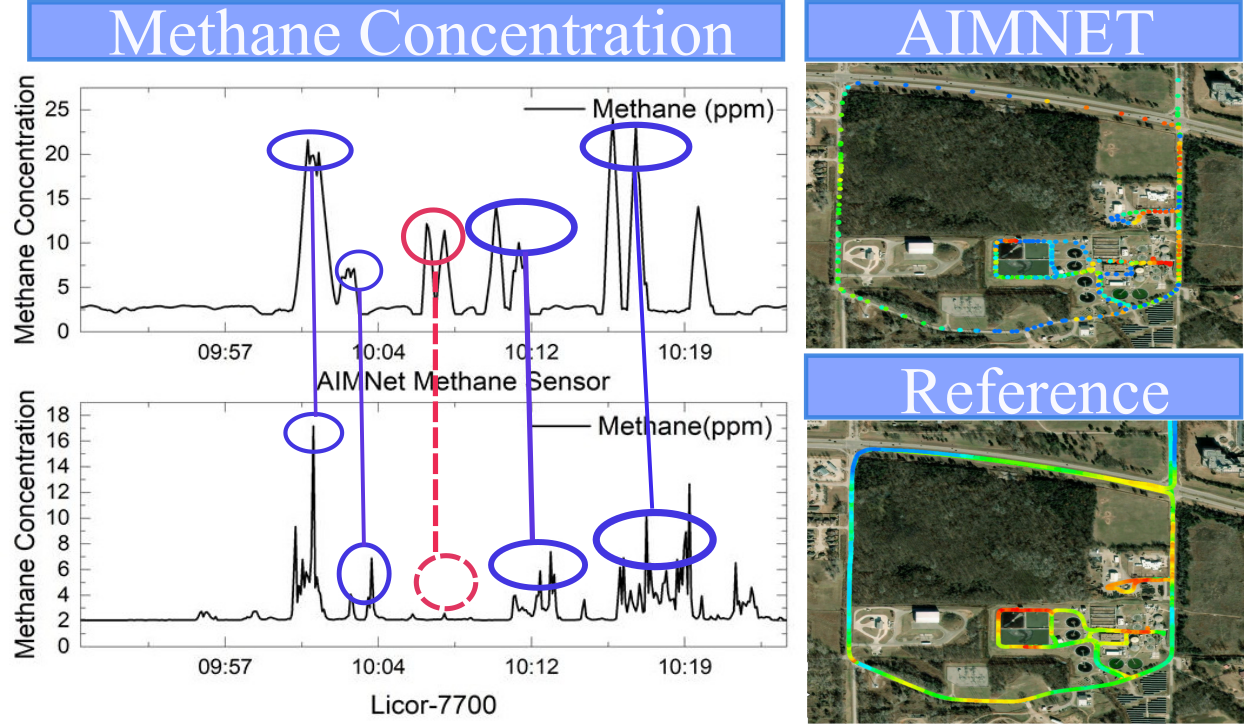}
    \caption{Mobile $\text{CH}_4$ measurement.}
    \label{fig:field}
\end{figure}

For the dynamic mode, a mobile methane monitoring system is tested around the City of Norman Water Reclamation Facility to demonstrate a practical IoT-based solution for detecting and modeling local and regional methane leakage.
To validate the accuracy of our device, we employ a high-precision LI-7700 sensor as the reference standard and conducted comparative measurements using~{\name}. Fig.~\ref{fig:field} presents the testing results. The upper two images show the output from~{\name}, while the lower image displays the reference measurements. It can be observed that~{\name} successfully captures all the major methane concentration peaks detected by the LI-7700. This indicates that when the methane concentration exceeds 5 ppm, our sensor is capable of accurately identifying the elevated levels. Only one false positive is reported, with a concentration reading of 15 ppm.
\textcolor{black}{We observe a time offset between them, attributable to their fundamentally different sensing architectures: the LI-7700 directly measures methane concentration in the ambient air stream with negligible residence time, whereas {\name} utilizes an active pumped-inlet designed for static environmental monitoring. This configuration introduces a small but systematic transport delay, particularly under highly dynamic conditions, since air must traverse the intake tubing and mixing chamber before reaching the sensing cell.}
The geographic information system-based figure on the right provides a detailed spatial visualization of methane detection locations. It further confirms that~{\name} successfully identifies all high-concentration methane areas that are also detected by the LI-7700. The only missed region is a mid-level concentration zone (approximately 4 ppm) located in the lower left corner. 
The 15 ppm false positive occurred on the highway, likely due to rapid vehicle motion. In detail, our mobile platform incorporates a vehicle-mounted system, a drone-equipped optical gas imaging camera, and an integrated weather station to collaboratively capture real-time methane plume dynamics. 

We further apply our multiscale modeling framework over the same study site, nesting the geometry-resolving LES model within the WRF-GHG simulation domain. To illustrate the value of explicitly resolving built structures, here we first increase the WRF-GHG resolution to 32 m, approaching LES-scale turbulence representation but without the ability to resolve building geometry. We then conduct a geometry-resolving LES simulation at a finer horizontal resolution of 5 m, using lateral boundary conditions from the 32-m WRF-GHG simulation.
\begin{figure}[htbp]
\centering
    \includegraphics[width=1.0\linewidth]{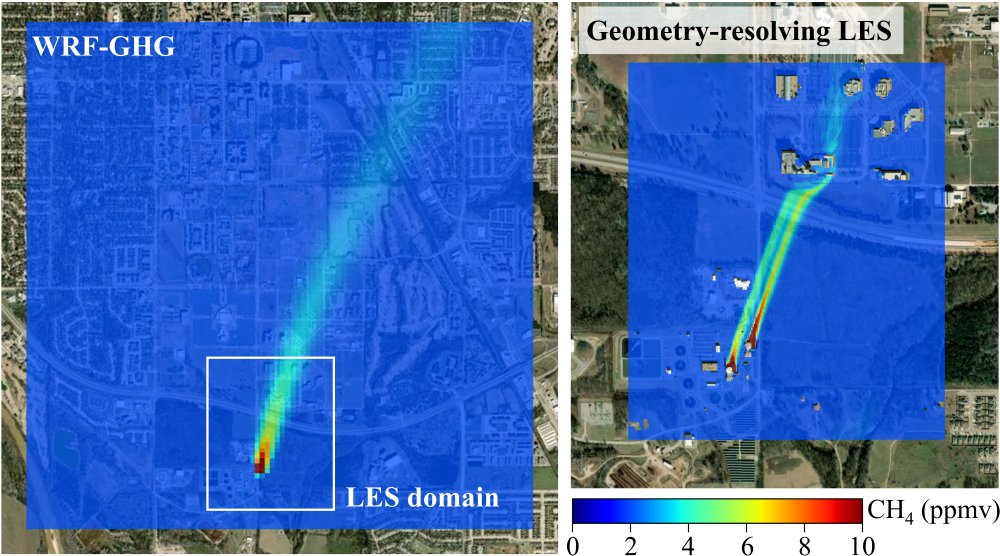}
    \caption{Physics-based multiscale modeling system.}
    \label{fig:model}
\end{figure}

Fig.~\ref{fig:model} compares the results from both simulations. While both models reasonably capture CH$_4$ concentrations near the emission sources (anaerobic digesters), substantial differences emerge farther downwind, particularly in regions behind buildings. The geometry-resolving LES successfully reproduces the meandering behavior of the CH$_4$ plume, a feature not captured by the coarser WRF-GHG simulation. In addition, the high-resolution geometry-resolving LES can spatially differentiate emissions from multiple, closely spaced sources that would otherwise be lumped together within a single WRF-GHG grid cell. Compared to WRF-GHG, the LES also aligns more closely with field measurements taken along the road east of the emission sources (Fig.~\ref{fig:field}), providing a more accurate representation of plume dispersion. This improved agreement is critical for the reliable identification and quantification of emission sources and their strength. Since WRF-GHG is typically run at even coarser resolutions in real-world, operational deployments (e.g., 1 km), these results highlight the value of a multiscale modeling system, particularly when applied to oil and gas infrastructure with complex geometries.

\section{Addressing Key Challenges with AIMNET}
\label{sec:Challenge}

Achieving fully autonomous and self-regulating carbon-based gas monitoring and hazard prediction remains challenging, particularly in large-scale deployment and accurate detection. Next, we outline the key research challenges and describe how~{\name} is specially designed to address them. 

\textbf{Reliable Sensing \& Edge Intelligence}: 

Reliable gas sensing is always a significant challenge for large-scale deployment. Low-cost sensors enable broad spatial coverage but suffer from limited sensitivity, signal drift, and cross-gas interference, where issues are exacerbated by environmental factors such as humidity and temperature that corrupt the data. For example, in oil fields or agricultural zones, CH$_4$ often coexists with CO$_2$ and volatile organic compounds, hindering its detection accuracy. Conversely, while high-precision instruments like tunable diode laser absorption spectroscopy and Fourier-transform infrared spectrometers offer superior accuracy, their cost, power requirements, and bulk render them impractical for the widespread, distributed networks essential for regional-scale monitoring.

{\name} directly addresses this dilemma through an innovative blend of custom hardware design and ML-driven calibration. The system continuously ingests real-time sensor streams and environmental metadata, applying ML-based correction mechanisms to account for sensor drift and environmental variability. This closed-loop framework enhances sensing reliability and eliminates the frequent manual recalibration. Moreover, leveraging historical data and predictive modeling, {\name} validates readings, reconstructs missing or corrupted values due to communication or sensor faults, and identifies anomalous gas patterns. These self-healing capabilities enable long-term autonomous operation and precise detection of leaks and hazardous events, making it ideal for scalable carbon-based gas monitoring in diverse and dynamic field settings.

\textbf{Scalable Real-Time Analytics \& Visualization}: %% 

Large-scale distributed sensors generate massive data streams, creating significant hurdles for scalable, real-time emission detection. High data throughput, coupled with inconsistent formats and asynchronous inputs, makes sub-second analytics a formidable challenge. Traditional models like Integrated Moving Average are ill-equipped to handle the nonlinear, non-stationary dynamics of gas dispersion, especially with noisy or incomplete data~\cite{khiabani2024challenges}. Furthermore, operators face cognitive overload from cluttered visualization interfaces that fail to convey spatial-temporal patterns and data uncertainty clearly.

{\name} addresses these challenges with a robust, intelligent analytics core that enables high-frequency monitoring and real-time visualization. Its modular, scalable architecture leverages distributed computing tools such as Apache Kafka and Spark for high-volume, sub-second data processing, with the potential for lower-latency stream-first integrations. To overcome the limitations of the conventional statistical models, {\name} employs advanced ML pipelines, like long short-term memory networks, for accurate gas dispersion modeling. The DT framework enables accurate, physics-informed leak localization by comparing real-time sensor data with LES outputs, dynamically integrating live environmental data, background gas concentrations, and weather conditions, thereby distinguishing true anomalies and reducing false positives. 
The system supports semantic temporal fusion for aligning asynchronous data streams. Insights are delivered through a web-based, interactive visualization platform, designed for future integration with Large Language Model-powered natural language queries (e.g., “What is the most likely emission source near node 17?”), transforming fragmented data into clear, actionable intelligence.

\textbf{Trustworthy \& Secure AI Integration}: 

Deploying AI in large-scale environmental monitoring holds transformative promise but also faces key challenges. Reliable models require extensive, high-quality labeled data, which is often scarce due to the rarity of critical events and noisy field data (e.g., methane leaks, sensor tampering). Real-time performance necessitates balancing inference accuracy and computational efficiency, particularly for edge computing. Moreover, in safety-critical and regulated settings, AI outputs must be transparent, explainable, and auditable.

To address these challenges, {\name} adopts a modular AI framework that leverages edge-cloud cooperation and is deeply integrated with the DT system. The DT framework provides historical context, environment-aware feedback, and data traceability visualization, all of which enhance model reliability and facilitate continuous performance evaluation. For secure and compliant AI deployment, {\name} incorporates two key mechanisms: application-layer encryption to protect data beyond transport-level security, and DT-Honeypot to proactively detect and analyze malicious behaviors. Finally, {\name} supports AI lifecycle management by enabling model retraining, recalibration, and concept drift detection over time, ensuring long-term reliability, trustworthiness, and compliance of AI-driven functions.

\section{Conclusion}
We developed~{\name}, an IoT-powered Digital Twin (DT) system designed for continuous gas emission monitoring and early hazard detection. {\name} transcends the limitations of current monitoring approaches, which are often geographically constrained and lack intelligence, by providing a scalable, resilient, and fully integrated solution. The three-layer architecture facilitates a seamless data flow from sensors to analytics.  \textcolor{black}{Key innovations of~{\name} include the development of a low-power, self-calibrating IoT sensing instrument, a stable bidirectional communication framework, and an AI-driven analytics core built on the digital twin world that enables predictive modeling and real-time decision support.} 
Then we verified~{\name} credibility based on actual deployment. By holistically addressing the challenges, {\name} serves as a concrete DT blueprint for the next generation of smart environmental monitoring systems.

\section*{Acknowledgements}
This work was supported by the DOE iM4 (Innovative Methane Measurement, Monitoring, and Mitigation) program under the contract number: DE-FE0032285, and NSF CPS-2331105.

The authors utilized AI-assisted language polishing tools to improve the readability of the manuscript. These tools were not used for generating scientific content; all conceptual development, technical content, analyses, and all scientific conclusions are entirely original.

\bibliographystyle{IEEEtran}
\bibliography{ref}

\end{document}